# Title: Photons that travel in free space slower than the speed of light

**Authors:** Daniel Giovannini[1†], Jacquiline Romero[1†], Václav Potoček[1], Gergely Ferenczi[1], Fiona Speirits[1], Stephen M. Barnett[1], Daniele Faccio[2], Miles J. Padgett[1*]

**Affiliations:**

[1] School of Physics and Astronomy, SUPA, University of Glasgow, Glasgow G12 8QQ, UK

[2] School of Engineering and Physical Sciences, SUPA, Heriot-Watt University, Edinburgh EH14 4AS, UK

[†] These authors contributed equally to this work.

* Correspondence to: miles.padgett@glasgow.ac.uk

**Abstract:** That the speed of light in free space is constant is a cornerstone of modern physics. However, light beams have finite transverse size, which leads to a modification of their wavevectors resulting in a change to their phase and group velocities. We study the group velocity of single photons by measuring a change in their arrival time that results from changing the beam's transverse spatial structure. Using time-correlated photon pairs we show a reduction of the group velocity of photons in both a Bessel beam and photons in a focused Gaussian beam. In both cases, the delay is several microns over a propagation distance of the order of 1 m. Our work highlights that, even in free space, the invariance of the speed of light only applies to plane waves. Introducing spatial structure to an optical beam, even for a single photon, reduces the group velocity of the light by a readily measurable amount.

**One sentence summary:** The group velocity of light in free space is reduced by controlling the transverse spatial structure of the light beam.

**Main text**

The speed of light is trivially given as $c/n$, where $c$ is the speed of light in free space and $n$ is the refractive index of the medium. It follows that in free space, where $n = 1$, the speed of light is simply $c$. We show experimentally that the introduction of transverse structure to the light beam reduces the group velocity by an amount depending upon the aperture of the optical system. The delay corresponding to this reduction in the group velocity can be many times greater than the optical wavelength and consequently should not be confused with the $\approx \pi$ Gouy phase shift (*1, 2*). To emphasize that this effect is both a linear and intrinsic property of light, we measure the delay as a function of the transverse spatial structure of single photons.

The slowing down of light that we observe in free space should also not be confused with slow, or indeed fast, light associated with propagation in highly nonlinear or structured materials (*3, 4*). Even in the absence of a medium, the modification of the speed of light has previously been known. For example, within a hollow waveguide, the wavevector along the guide is reduced below the free-space value, leading to a phase velocity $v_\phi$ greater than $c$. Within the hollow waveguide, the product of the phase and group velocities is given as $v_\phi v_{g,z} = c^2$, thereby resulting in a group velocity $v_{g,z}$ along the waveguide less than $c$ (5).

Although this relation for group and phase velocities is derived for the case of a hollow waveguide, the waveguide material properties are irrelevant. It is simply the transverse spatial confinement of the field that leads to a modification of the axial component of the wavevector, $k_z$. In general, for light of wavelength $\lambda$, the magnitude of the wavevector, $k_0 = 2\pi/\lambda$, and its Cartesian components $\{k_x, k_y, k_z\}$ are related through (5)

$$k_z^2 + k_x^2 + k_y^2 = k_0^2.$$

As all optical modes of finite $x, y$ spatial extent require $k_x, k_y > 0$, which implies $k_z < k_0$, giving a corresponding modification of both the phase and group velocities of the light. Note that, for a fixed value of $k_x$ and $k_y$, $k_z$ is dispersive even in free space.

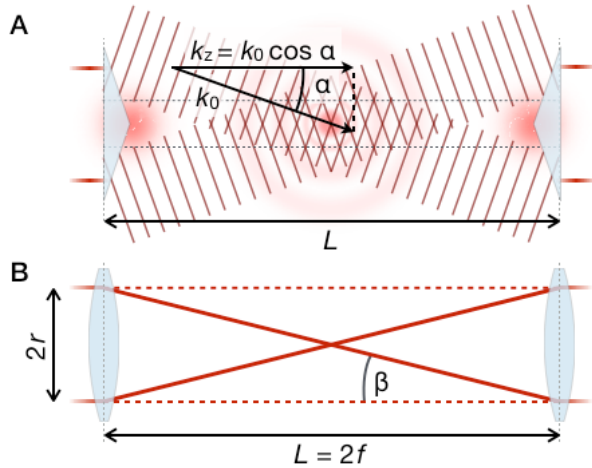

**Figure 1.** (**A**) A Bessel beam can be created using an axicon producing conical phase fronts of angle α. (**B**) A ray entering a confocal telescope at radius $r$ will travel an additional distance proportional to $\cos^{-1} \beta$.

Extending upon the case of a mode within a hollow waveguide, an example of a structured beam is a Bessel beam (Fig. 1A), which is itself the description of a mode within a circular waveguide (1, 6). In free space, Bessel beams can be created using an axicon, or its diffractive optical equivalent (7), that converts a plane wave into conical phase fronts characterized by a single radial component of the wavevector, $k_r = \sqrt{k_x^2 + k_y^2}$ (8–10). This single value of the radial component gives a unique value of $k_z < k_0$ and hence uniquely defined phase and group velocities (11).

In our work, we want to avoid complications arising from the finite thickness of refractive optical elements. We therefore use diffractive optics, idealized as having zero thickness. For a Bessel beam created with a diffractive optic (7), characterized by $k_r$ (with $k_r \ll k_0$), the axial component of the wavevector is given by $k_z = k_0 - k_r^2/2k_0$. The resulting phase velocity and group velocity along $z$ are

$$v_\phi = c\left(1 - \frac{k_r^2}{2k_0^2}\right)^{-1} \text{ and}$$

$$v_{g,z} = c\left(1 - \frac{k_r^2}{2k_0^2}\right).$$

This modification of the phase and group velocities of Bessel beams has been examined in the classical, many-photon regime. Subtle changes in velocity have been previously studied using Bessel beams in the microwave (*12*) and optical regimes (*13–15*).

We demonstrate the intrinsic, and linear, nature of this reduction in group velocity by measuring the delay in the arrival time of single photons. Over a propagation distance of $L$, the reduction in the group velocity compared to the plane-wave case gives a delay of

$$\delta z_{\text{Bessel}} \approx L\frac{k_r^2}{2k_0^2} = \frac{L}{2}\alpha^2. \quad (1)$$

As an example, for an axicon designed to produce $\alpha = k_r/k_0 = 4.5 \times 10^{-3}$ over a propagation distance of 1 m, we predict a delay of ~30 fs, corresponding to a spatial delay of 10 µm.

Measuring the arrival time of single photons with femtosecond precision is challenging. Consequently, we adopt a method relying upon a quantum effect, namely, the Hong-Ou-Mandel (HOM) interference (*16*). We use a parametric down-conversion source to produce photon pairs that are very strongly correlated in their wavelengths and their generation time. One photon can then act as a reference, against which the arrival of the other photon can be compared. When the arrival times of the two photons incident on a beam splitter are matched to a precision better than their coherence time, both photons emerge from the same output port. Under this matched condition, the coincidence rate for detection at the two output ports of the beam splitter falls to zero, which results in what is known as a Hong-Ou-Mandel dip.

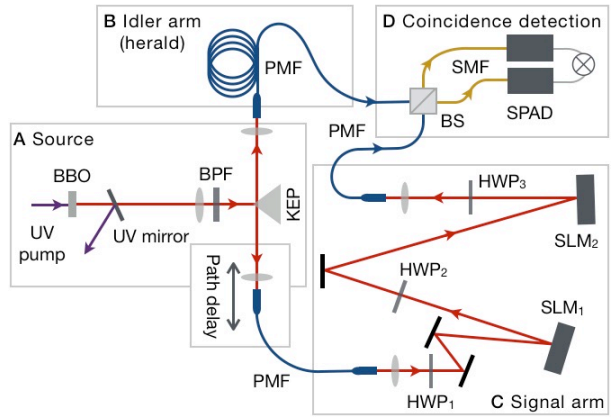

**Figure 2.** Experimental apparatus. Photon pairs from a parametric down-conversion source are separated by a knife-edge prism (KEP); a band-pass filter (BPF) sets the spectral profile of the down-converted light. Half-wave plates (HWP) are used to maximize the efficiency of the spatial light modulators (SLM), and match the polarization of the polarization-maintaining fibers (PMF). Signal and idler photons interfere at a beam splitter (BS) whose outputs are coupled to single-mode fiber (SMF), connected to avalanche photodiodes (SPAD). The SPADs feed a coincidence counter. The path delay of the signal photon is adjusted by means of a translation stage, and the position of the HOM dip recorded as a function of the spatial shaping of the photon.

We use an ultraviolet laser incident upon a beta-barium borate (BBO) crystal to produce photon pairs with central wavelength at 710 nm. The photons, called signal and idler, pass through an interference filter of spectral bandwidth 10 nm and are collected by polarization-maintaining, single-mode fibers. One fiber is mounted on an axial translation stage to control the path length (Fig. 2A). The idler photon goes through polarization-maintaining fibers before being fed to the input port of a fiber-coupled beam splitter (Fig. 2B) (*17*). Instead of going straight to the other beam splitter input, the signal photon is propagated through a free-space section (Fig. 2C). This consists of fiber-coupling optics to collimate the light and two spatial light modulators (SLMs). SLMs are pixelated, liquid-crystal devices that can be encoded to act as diffractive optical elements implementing axicons, lenses and similar optical components. The first SLM can be programmed to act as a simple diffraction grating such that the light remains collimated in the intervening space, or programmed to act as an element to structure the beam (e.g. axicons or lenses with focal length $f$). The second SLM, placed at a distance $2f$, reverses this structuring so that the light can be coupled back into the single-mode fiber that feeds to the other input port of the beam splitter. The output ports of the fiber-coupled beam splitter are connected to single-photon detectors, which in turn feed a gated counter (Fig. 2D). The coincident count rate is then recorded as a function of path difference between the signal and idler arms. The position of the HOM dip is recorded as a function of the spatial shaping of the signal photon.

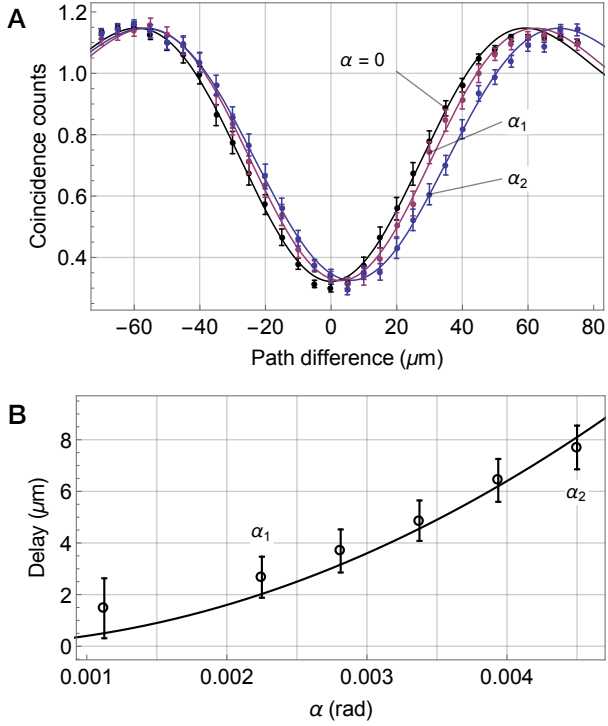

**Figure 3.** Experimental results for a Bessel beam. (**A**) Measured Hong-Ou-Mandel dips for two values of α ($α_1$ = 0.00225, red; $α_2$ = 0.00450, blue) and the α = 0 case (black), with corresponding best fits (*16*). (**B**) Measured delays (hollow circles) and theoretical prediction from Eq. 1 (solid line), for different values of α. The delays are expressed with respect to the α = 0 case, corresponding to an unstructured collimated beam.

Taking the Bessel beam as our first example, the transverse structuring can be turned on and off for each value of path difference. The corresponding position of the HOM dip can then be directly compared between the two cases. Figure 3A shows the baseline-normalized coincidences for two different values of $α = k_r/k_0$ (where we define the baseline as the coincidence count at path differences far from the dip position). We note that in all cases the width of the HOM dips is the same, set by the 10 nm spectral bandwidth of the down-converted photons. The key result is that the HOM dip associated with the Bessel beam is delayed with respect to the dip obtained for a collimated beam. We measure a delay of 2.7±0.8 µm for the case of $α_1$ = 0.00225 rad and 7.7±0.8 µm for $α_2$ = 0.00450 rad. These measured values agree with theoretical predictions of 2.0 µm and 8.1 µm for $α_1$ and $α_2$, respectively.

The analytical form of this predicted delay (Eq. 1) suggests a simple geometrical model, where the delay arises from the additional length of the diagonal ray, propagating at an angle $α$ with respect to the optical axis. In Fig. 3B we compare the measured and predicted values for the delay, showing that Eq. 1 is valid over the range of Bessel angles that we tested.

Perhaps the most common form of spatial structuring of a light beam is focusing, which also leads to a modification of the axial component of the wavevector. We consider the propagation of light through a telescope comprising two identical lenses separated by twice their focal length, $f$ (i.e. a confocal telescope). Assuming a ray-optical model, a co-axial ray incident upon the first lens at radius $r$ emerges from the second lens co-axially at the same radius but inverted about the optical axis (Fig. 1B). Comparing the on-axis separation of the lenses to this diagonal distance gives an additional distance traveled of $\delta z = L/\cos \beta - L \approx r^2/f$, where $\beta$ is the angle between ray and optical axis.

For a beam of Gaussian intensity distribution with $1/e^2$ radius $w$, the expectation value of $r^2$ is $\langle r^2 \rangle = w^2/2$. Therefore, the expected delay $\delta z_{\text{Gauss}}$ for a Gaussian beam on transmission through a confocal telescope is

$$\delta z_{\text{Gauss}} = w^2/2f = (w/f)^2 \times f/2, (2)$$

where $w$ is the waist of the input beam. The delay is a quadratic function of the quantity $w/f$, which can be considered as a measure of the beam divergence, defined by the numerical aperture of the system. The delay increases with increasing numerical aperture. This geometrical model and a rigorous theoretical calculation provide the same results for both the Bessel and confocal cases, within the same approximations (*18*). The full theoretical model, however, applies to any arbitrary field. As the delay increases with the square of the numerical aperture, the delay becomes progressively harder to detect at longer distances.

The delay arising from focusing is shown in Fig. 4. Trace A shows the position of the HOM dip for the case of a collimated beam, and trace B shows its position for the case of $f = 0.40$ m. We measure a delay of 7.7±0.4 µm for the focused case. This is comparable to the predicted delay based on Eq. 2 which, for our beam of $w = 2.32 \pm 0.09$ mm, is 6.7± 6 µm. The slight difference between our measurement and the predicted value is likely due to residual aberrations and imperfect collimation, leading to an ill-defined beam waist, upon which the delay is quadratically dependent.

We further investigate the dependence of the delay upon the beam structure by introducing aperture restrictions to the beam, in the form of center and edge stops (insets of Fig. 4). Results are shown in traces C and D in Fig. 4, together with the full-aperture focused beam case (red line, trace B). A center stop increases the expectation value of $r^2$, thereby increasing the delay compared to the full-aperture case. Trace C shows the dip with a center stop of radius 1.4 mm, as shown in inset C. We measure a dip position additionally delayed by 7.3±0.4 µm compared to the full-aperture focused beam, giving a total delay of 15.0±0.6 µm. Next, we introduce an edge stop of the same radius, as shown in inset D. By restricting the aperture, the expectation value of $r^2$ is decreased, decreasing the delay with respect to the collimated case. Trace D shows the position of the HOM dip, which is now reduced by 6.4±0.4 µm with respect to the full-aperture case, resulting in a total delay compared to the collimated case of 1.3±0.6 µm.

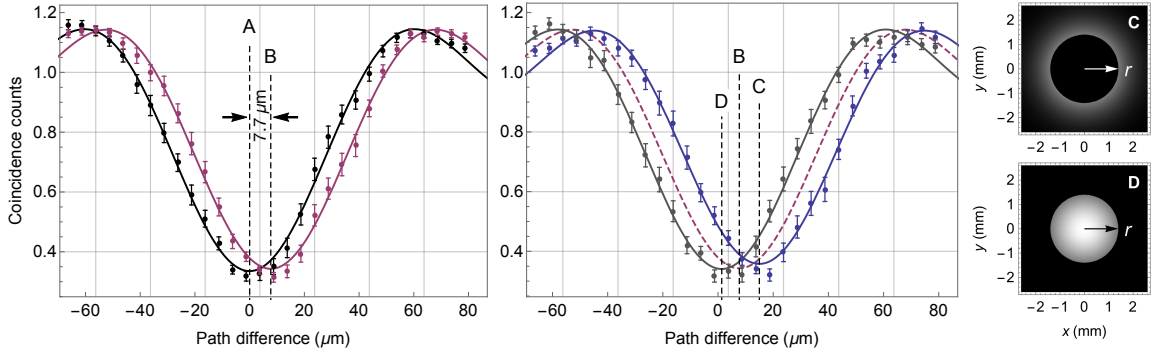

**Figure 4.** Measured Hong-Ou-Mandel (HOM) dips for collimated and focused Gaussian beams. (**Left panel**) HOM dip comparison for collimated (black) and focused (red) Gaussian beam. Minima are marked by A and B. (**Right panel**) HOM dip comparison for cases with an $r = 1.4$ mm center stop (blue, corresponding to inset C), and an edge stop of the same radius (gray, corresponding to inset D). Minima are marked by C and D. The red dashed curve is shown as reference. For each path difference, the counts for the four SLM settings were obtained consecutively.

It is important to consider three possible sources of systematic errors. Firstly, the phase values of all the pixels of the SLMs lie between 0 and $2\pi$ with an average value of $\approx \pi$. Regardless of what optical component is encoded on the SLMs, the effective thickness of the liquid crystal, as averaged over the full aperture, remains the same. Consequently, the observed delay is not a result of the SLMs themselves. Secondly, the width of the HOM dip remains compatible with the interference filter used. Therefore the coherence time of the light is unchanged by the setting of the SLMs and therefore the magnitude of the delays cannot be a result of spectral post-selection. Thirdly, one must ensure that the delays are not due to misalignment in the optical paths. In aligning the experiment, we employed back-projection following the Klyshko picture (*19*). More importantly, the alignment for the cases where we have aperture restrictions remains the same (the coaxial apertures do not change the path of the beam). Hence, the delays we measure can only result from the transverse structure of the beam and indeed are consistent with our theoretical predictions.

The speed of light in free space propagation is a fundamental quantity. It holds a pivotal role in the foundations of relativity and field theory, as well as in technological applications such as time-of-flight measurements, and radio and satellite communication. It has previously been experimentally established that single photons travel at the group velocity (*20*). We have now shown that transverse structuring of the photon results in a decrease in the group velocity along the axis of propagation. The effect can be derived from a simple geometric argument, which is also supported by a rigorous calculation of the harmonic average of the group velocity. Beyond light, the effect observed will have applications to any wave theory, including sound waves and, potentially, gravitational waves.

**Acknowledgements**

We thank Sonja Franke-Arnold for useful discussions. This work was supported by EPSRC through the COAM program, and by ERC through the TWISTS grant. We thank Hamamatsu for their support. M.J.P. thanks the Royal Society.

M.J.P., D.F. and S.M.B. conceived the experiment and supervised the work; M.J.P., D.G. and J.R. designed the experiment; D.G. and J.R. constructed the experiment, acquired the data and carried out the data analysis; V.P., J.R., G.F., F.S. and S.M.B. further developed the theoretical model; J.R., D.G. and M.J.P. wrote the text with input from all co-authors.